\documentclass[%
 reprint,
 superscriptaddress,
 amsmath,amssymb,
 aps,
 floatfix,
]{revtex4-1}

\usepackage[pdftex]{graphicx}
\usepackage{float}
\usepackage{epstopdf}
\usepackage{subfigure}
\usepackage{amsmath}
\usepackage{dcolumn}
\usepackage{bm}
\usepackage{verbatim}
\usepackage[colorlinks,
linkcolor=blue, urlcolor=blue, anchorcolor=blue, citecolor=blue]{hyperref}

\begin{document}

\preprint{APS/123-QED}

\title{Correlations in the chain melting of host-guest Calcium}

\author{Jiahao Dong}
\thanks{These two authors contributed equally}
\affiliation{
School of Physics, Nanjing University, Nanjing 210093, China}

\author{Yao Luo}
\thanks{These two authors contributed equally}
\affiliation{
School of Physics, Nanjing University, Nanjing 210093, China}

\author{Yong Wang}
\affiliation{
School of Physics, Nanjing University, Nanjing 210093, China}

\author{Jian Sun}
\email{jiansun@nju.edu.cn}
\affiliation{
School of Physics, Nanjing University, Nanjing 210093, China}

\date{\today}

\begin{abstract}

In this paper, we study chain melting state in the commensurate host-guest structure of calcium. We trained a NN (neural network) force field with DeepMD-kit, and by implementing classical simulations, we found that the convergence of z-proejcted MSD of guest atoms needs at least 8192-atom supercell. We ascribe the origination of chain melting to the phase transition at 600 K, in which the arrangement of guest chains transform from a checkerboard pattern to a stripe pattern, and the energy barrier of sliding chains against each other is obviously reduced. We also found that the chain melting state of calcium could be separated into two 2D (two-dimensional) and 3D (three-dimensional) disorder, with distinct chain dynamics.

\end{abstract}

\pacs{Valid PACS appear here}

\maketitle

\section{INTRODUCTION}

In recent years, the chain melting state in HG (host-guest) crystal structures has attracted lots of attention. Sodium~\cite{PRB-Single-crystal-studies-of-incommensurate-Na-to-1.5-Mbar}, potassium~\cite{PRB-Composite-incommensurate-K-III-and-a-commensurate-form-Study-of-a-high-pressure-phase-of-potassium}, rubidium~\cite{PRL-Pressure-Dependent-Incommensuration-in-Rb-IV}, calcium~\cite{PRL-Ca-A-Chain-Ordered-Host-Guest-Structure-of-Calcium-above-210-GPa}, scandium~\cite{PRB-Incommensurate-composite-crystal-structure-of-scandium-II} and barium~\cite{PRL-Self-Hosting-Incommensurate-Structure-of-Barium-IV} are usually converted to HG structures under high pressure. A typical HG structure can be separated into two sublattices: the guest sublattice, which is composed of 1D atomic chains, is located within the zeolite-shaped host sublattice. In such structures, diffraction peaks of the guest structure disappear upon heating, while those of the host structure remain, leading to a new phase: chain melting~\cite{PRL-Chain-Melting-in-the-Composite-Rb-IV-Structure,PRB-One-dimensional-chain-melting-in-incommensurate-potassium}.

It is important to figure out how the atoms of the guest sublattice move in the chain melting phase. One possibility is a 2D disorder state, in which the guest chains slide freely against each other, but the atoms within a chain move in a coherent manner; another possibility is a 3D disorder state, in which guest atoms within one chain diffuse, and inter-chain correlations are also lost. Neutron-scattering experiments show that the chain melting phase of incommensurate linear-chain mercury compound $Hg_{3-\delta}AsF6$ is consistent with a 2D disorder state. At room temperature, it has long-range order within one $Hg$ chain, but quite weak correlations between Hg chains ~\cite{PRL-One-Dimensional-Phonons-and-Phase-Ordering-Phase-Transition-in,PRL-One-Dimensional-Fluctuations-and-the-Chain-Ordering-Transformation-in}. MLMD (machine-learned molecular dynamics) simulations indicate that the chain melting phase of K-III (incommensurate HG structure of potassium) is chain 3D disorder state~\cite{PNAS-On-the-chain-melted-phase-of-matter}, in which intra-chain and inter-chain correlations vanish simultaneously. In incommensurate HG structures, the guest sublattice has a lattice constant incommensurate with the host unit cell, lacking periodicity in spite of possessing long-range order. Furthermore, guest chains are usually weakly coupled to each
other, leading to the formation of sliding phonon modes, in which guest chains can slide freely without energy penalty~\cite{SA-Strong-coupling-superconductivity-in-a-quasiperiodic-host-guest-structure,PRL-One-Dimensional-Phonons-and-Phase-Ordering-Phase-Transition-in}.

At 231 GPa, Ca-VII has been characterized as a host-guest structure (space group is $P4_2\/ncm$) with a commensurate host-guest ratio of 4/3~\cite{PRL-Ca-A-Chain-Ordered-Host-Guest-Structure-of-Calcium-above-210-GPa}. Instead of forming a straight line, the guest chain exhibits an s-shaped pattern along c axis, which increases the energy barrier of sliding chains relative to the host sublattice.

The development of NN (neural network) potential enables us to construct ab-initio quality force field, so that we can implement long-time scale MD simulations. It provides an accurate and efficient way to explore long-range correlation and long-time thermodynamic stability. 

Here, we trained DeepMD potential and perform classical molecular dynamics simulations to explore properties of chain melting phase of Ca-VII. We demonstrate that it successively undergoes three phase transitions: firstly, the ordered arrangement of guest chains transits from a checkerboard pattern to a stripe pattern at 600 K; upon heating, guest chains start to move in a manner much like 2D disorder; eventually both intra-chain and inter-chain correlations of guest atoms disappear, and the guest sublattice becomes 3D disorder.

\section{METHODS}\label{methods}

We performed AIMD (ab initio molecular dynamics) simulations with the VASP (Vienna Ab initio simulation package) code~\cite{PRB-Ab-initio-molecular-dynamics-for-liquid-metals,CMS-Efficiency-of-ab-initio-total-energy-calculations-for-metals-and-semiconductors-using-a-plane-wave-basis-set} and DFT (density-functional theory) with the PBE (Perdew-Burke-Ernzerhof) exchange-correlation functional~\cite{PRL-Generalized-Gradient-Approximation-Made-Simple}. The projector augmented-wave pseudopotential contains ten valence electrons, and the plane wave cutoff energy is fixed to 380 eV. AIMD simulations in the NVT (canonical) ensemble use Nosé-Hoover thermostats with a time constant of 200 fs and the time step is 1 fs. Simulation cells containing 128 atoms are used in all cases, for which the gamma point is used for Brillouin zone integration. We have also performed a few tests using LDA (local density approximation), different k-point values and cell sizes, and time step values as short as 0.25 fs. The results didn't show any significant variation. 

Classical MD simulations are performed using LAMMPS~\cite{JCP-Fast-Parallel-Algorithms-for-Short-Range-Molecular-Dynamics} code. We employed a Velocity Verlet integrator with a timestep of 1 fs. We performed isothermal simulations using the Stochastic Velocity Rescaling Thermostat~\cite{JCP-Canonical-sampling-through-velocity-rescaling}. We set the relaxation times for the thermostat at 100 fs. For simulations with the DeePMD potential, system sizes of 128, 512, 4096, 8192 and 18432 are used.

\begin{figure}[t]
\centering

\includegraphics[width=0.46\textwidth]{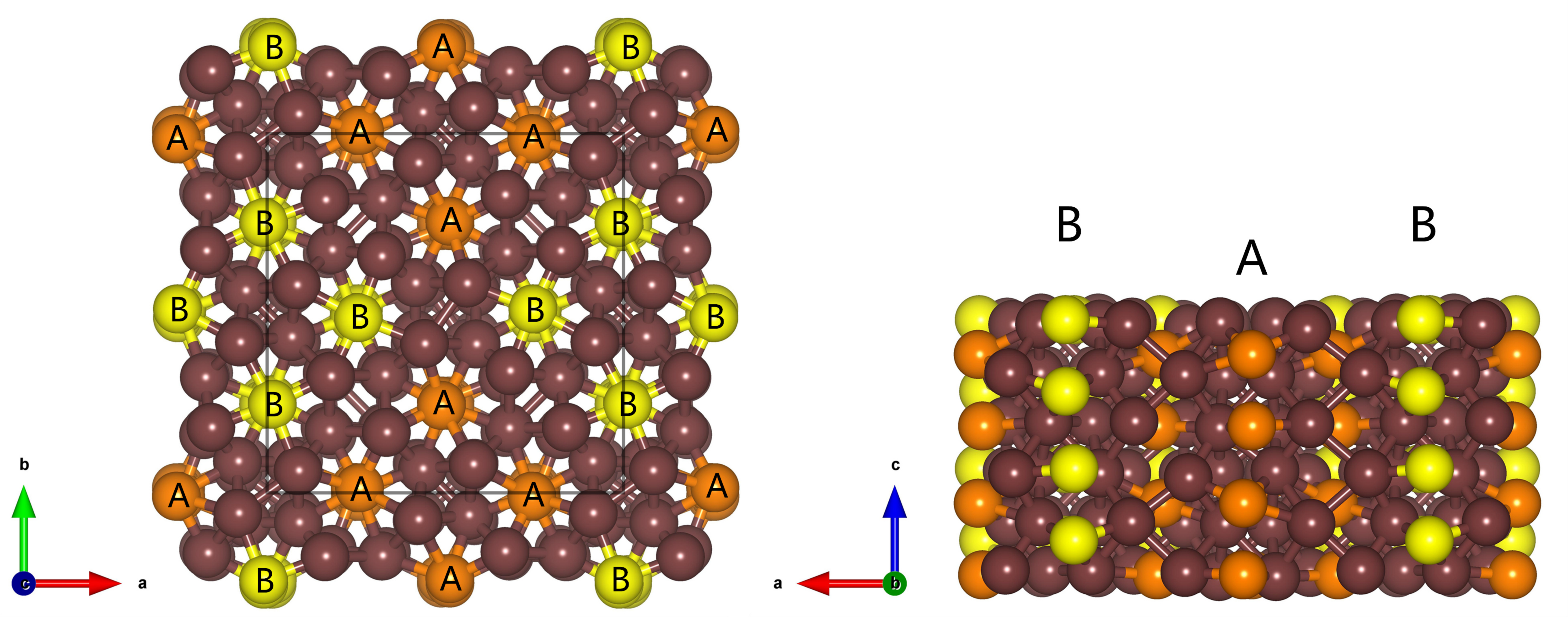}
\includegraphics[width=0.46\textwidth]{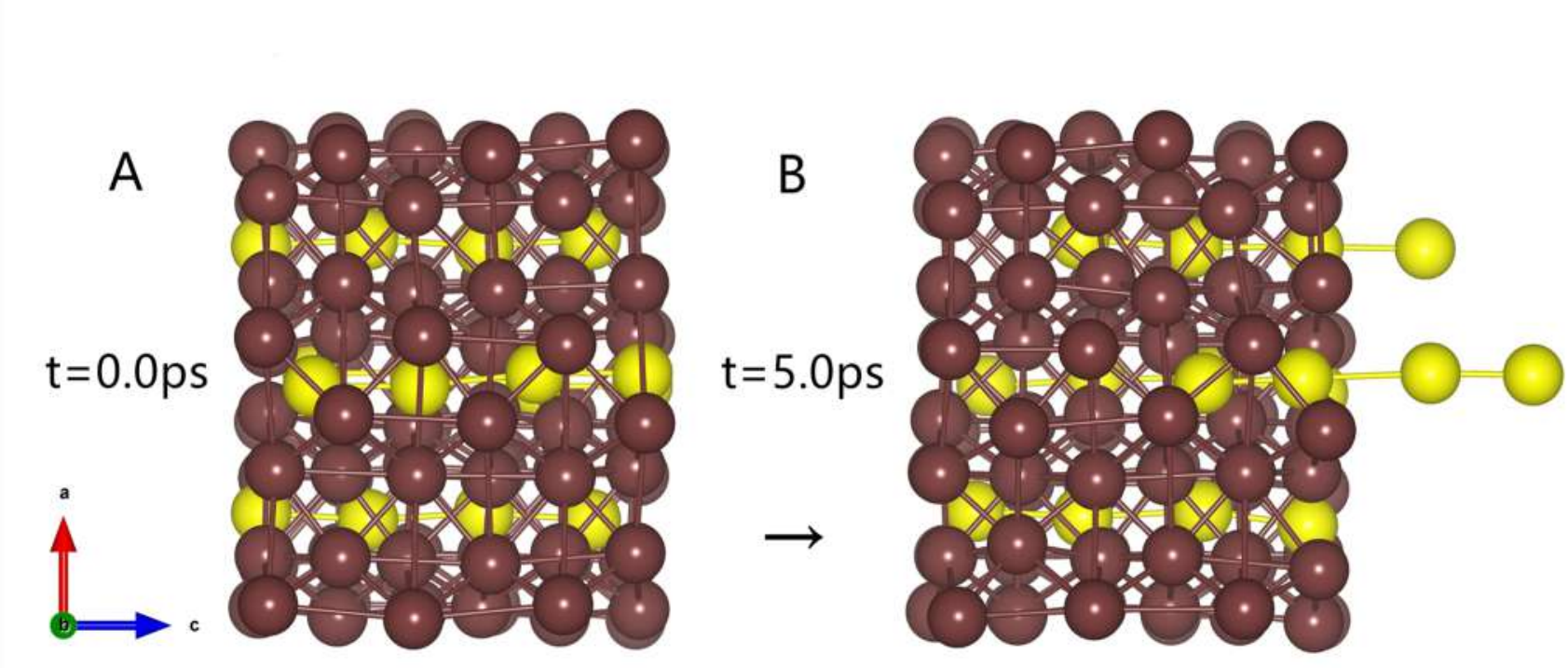}
\includegraphics[width=0.46\textwidth]{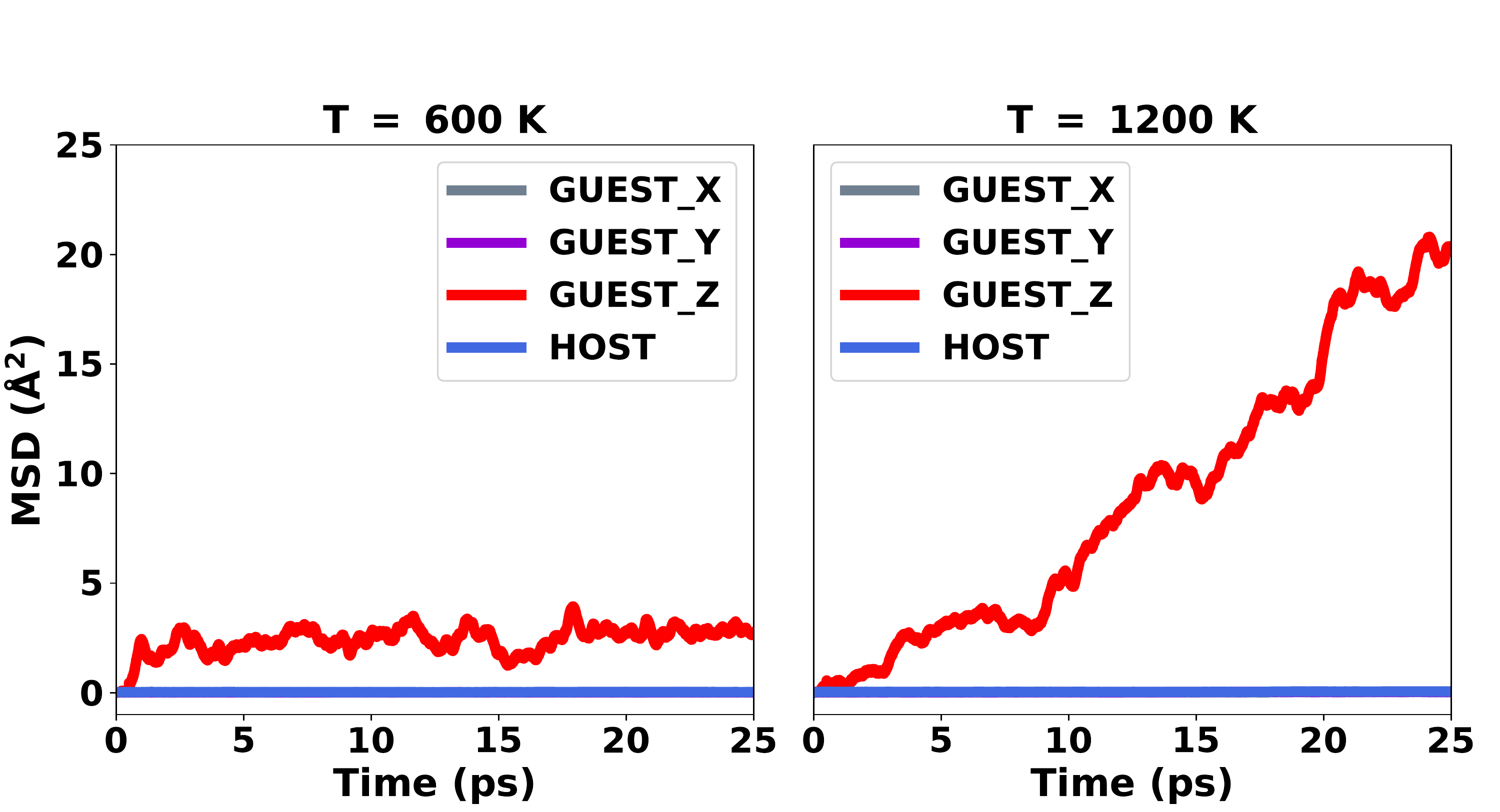}

\caption{(a) The crystal structure of Ca-VII drawn in the ab (left) and ac (right) planes. The labels A and B show the two different sublattices of the guest atoms. (b) Side view of Ca-VII at 231 GPa and T = 1200 K. (left) The initial structure for AIMD simulations. (right) The snapshot after 5.0 ps. (c) MSD (mean-square displacement) of guest and host atoms in Ca-VII in an 128-atom AIMD-NVT simulation at T = 600 K and 1200 K, respectively, and the pressure is P = 231 GPa.}
\label{vasp}
\end{figure}

\begin{figure}[t]
\centering

\includegraphics[width=0.46\textwidth]{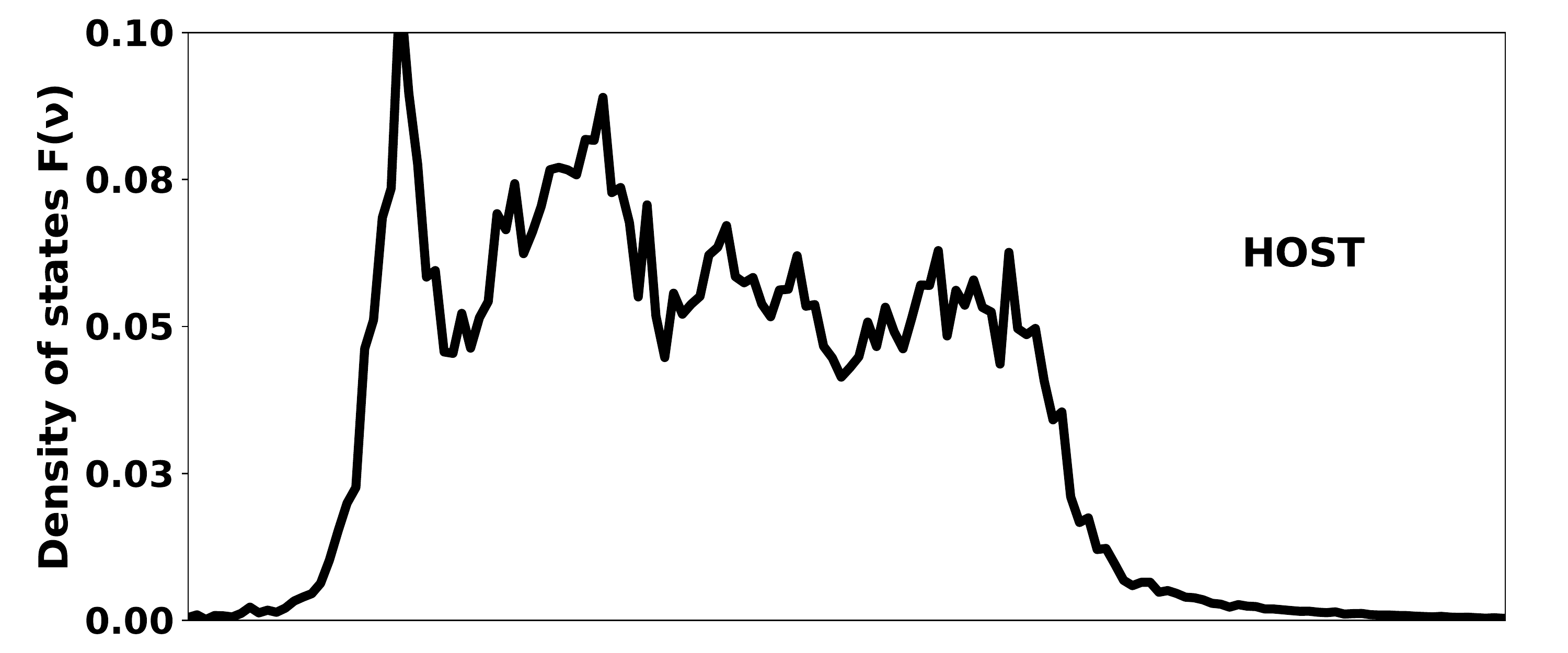}
\includegraphics[width=0.46\textwidth]{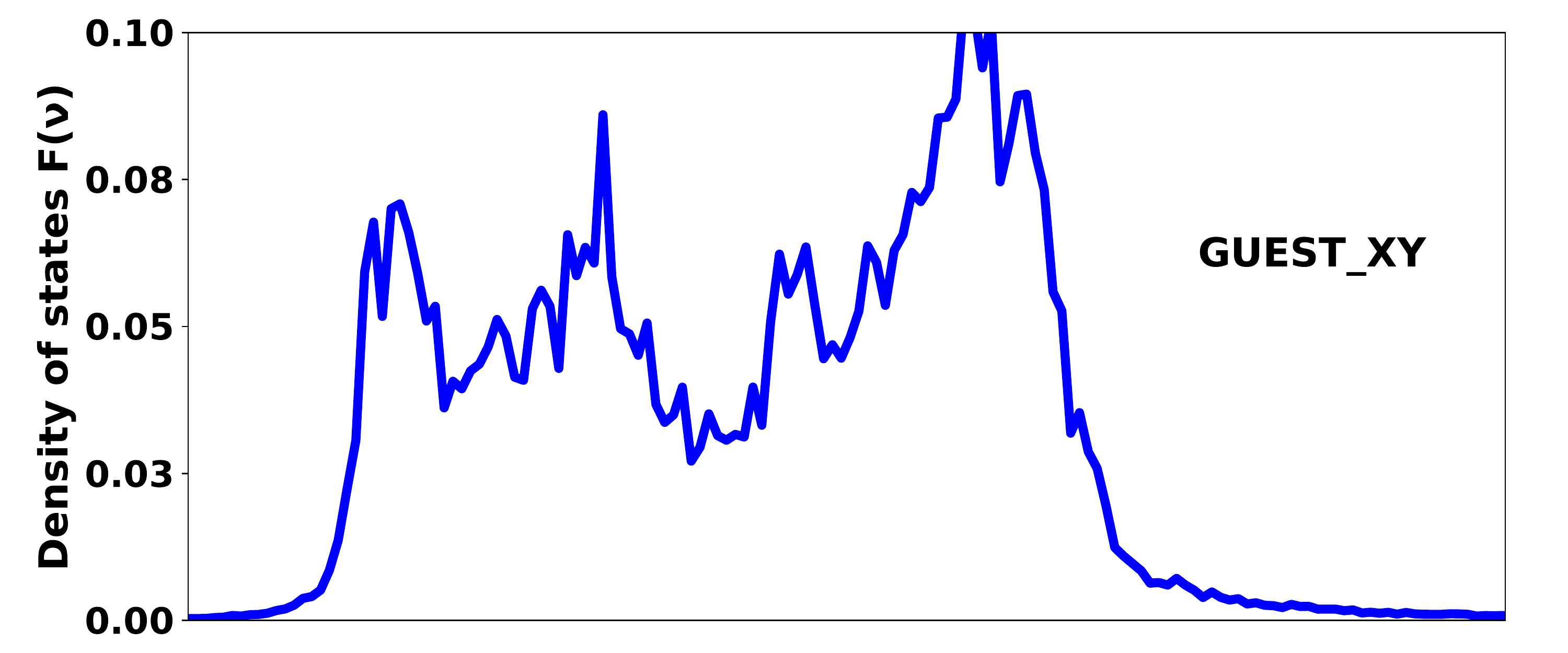}
\includegraphics[width=0.46\textwidth]{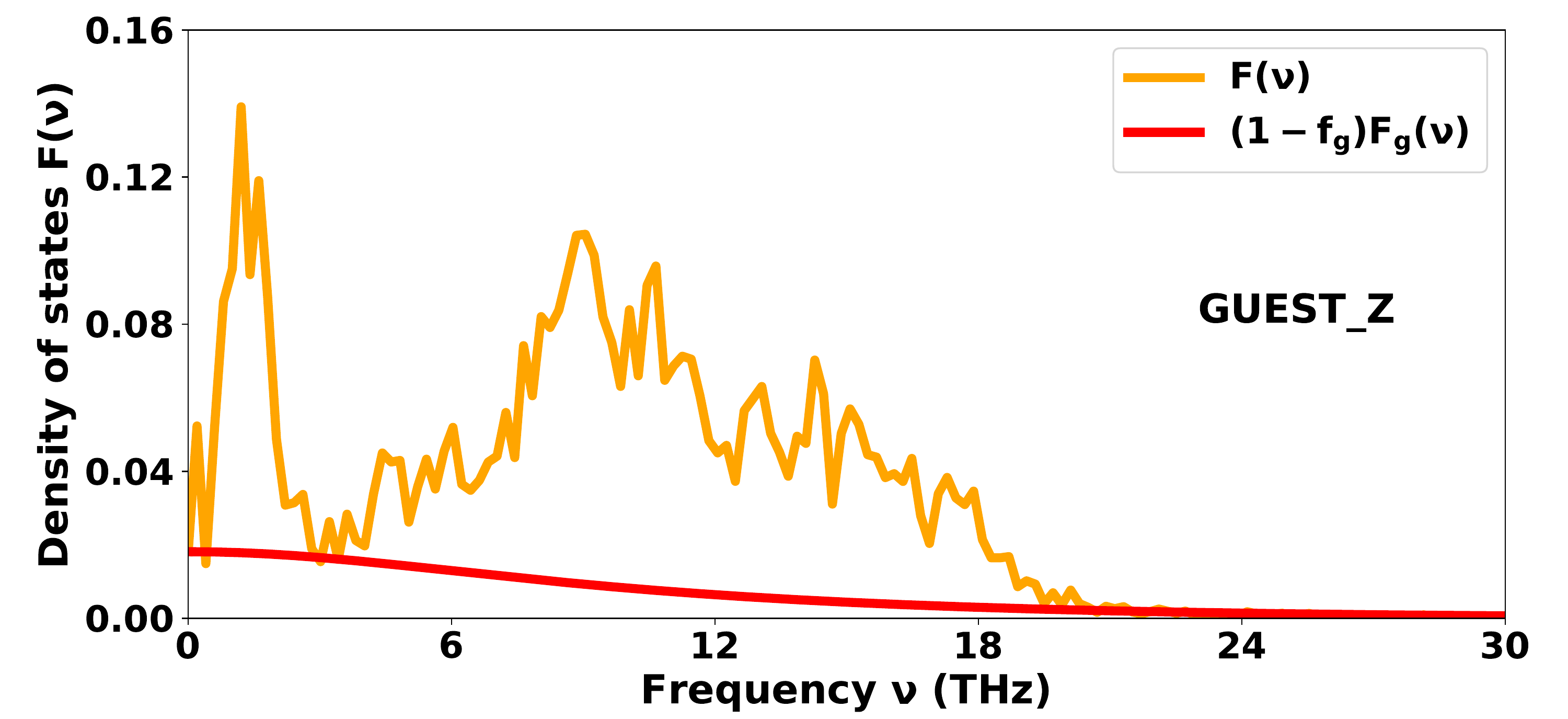}

\caption{Phonon density of states, which is obtained from Fourier transformation of velocity auto-correlation function, at 231 GPa and T = 1200 K.}
\label{vasp_spectrum}
\end{figure}

We use the DeePMD (deep potential molecular dynamics) scheme~\cite{PRL-Deep-Potential-Molecular-Dynamics-A-Scalable-Model-with-the-Accuracy-of-Quantum-Mechanics,CCP-Deep-Potential,CCP-Deep-Potential} to construct the force field. In DeepMD scheme, NNs (neural networks), which can handle complex quantum-mechanical data, are used to represent the potential energy surface. DeePMD builds a local reference frame for each atom, in order to construct invariant descriptors under all natural symmetries. We use DeePMD-kit~\cite{CPC-DeePMD-kit-A-deep-learning-package-for-many-body-potential-energy-representation-and-molecular-dynamics} to train the NN (neural network) potential and to interface the potential to LAMMPS. The training set includes 30,000 configurations, composed of a dataset of 24,000 snapshots extracted from equilibrium AIMD trajectories, ranging from 200 K to 3000 K, together with a dataset of 6,000 conﬁgurations from zero-temperature structures with randomly displaced atoms. We use a neural network with five hidden layers, and the number of neurons in each layer is equal to 240, 120, 60, 30, and 10, respectively. We trained the neural network for 800 epochs with ADAM optimizer, and we set an exponentially decaying learning rate changing from 1e-3 to 3.5e-8. During the optimization process, the prefactors of energy and the force terms in loss functions go from 0.02 to 8 and from 1000 to 1, respectively.

\section{RESULTS}

\begin{figure}[t]
\centering

\includegraphics[width=0.46\textwidth]{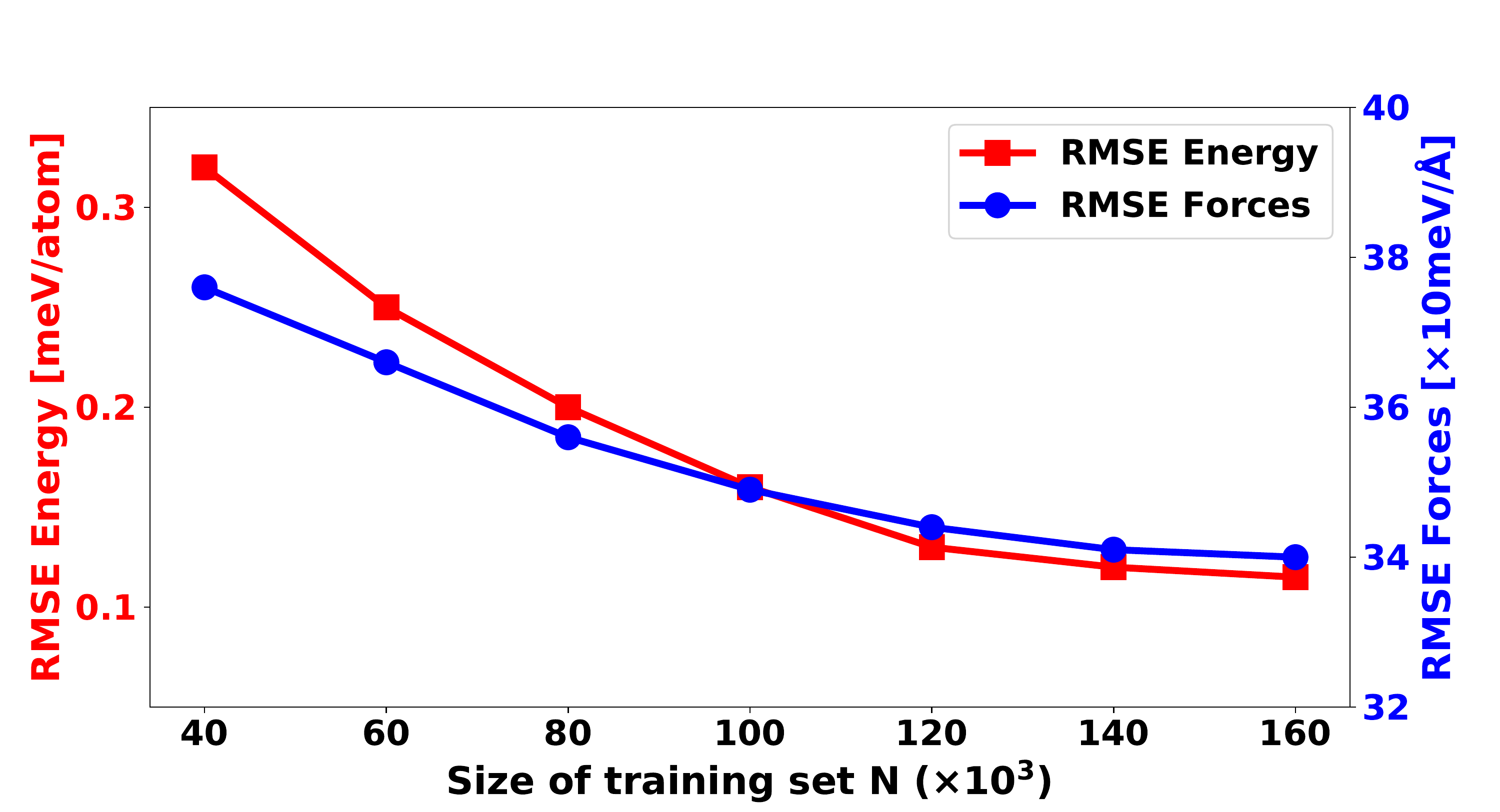}
\includegraphics[width=0.46\textwidth]{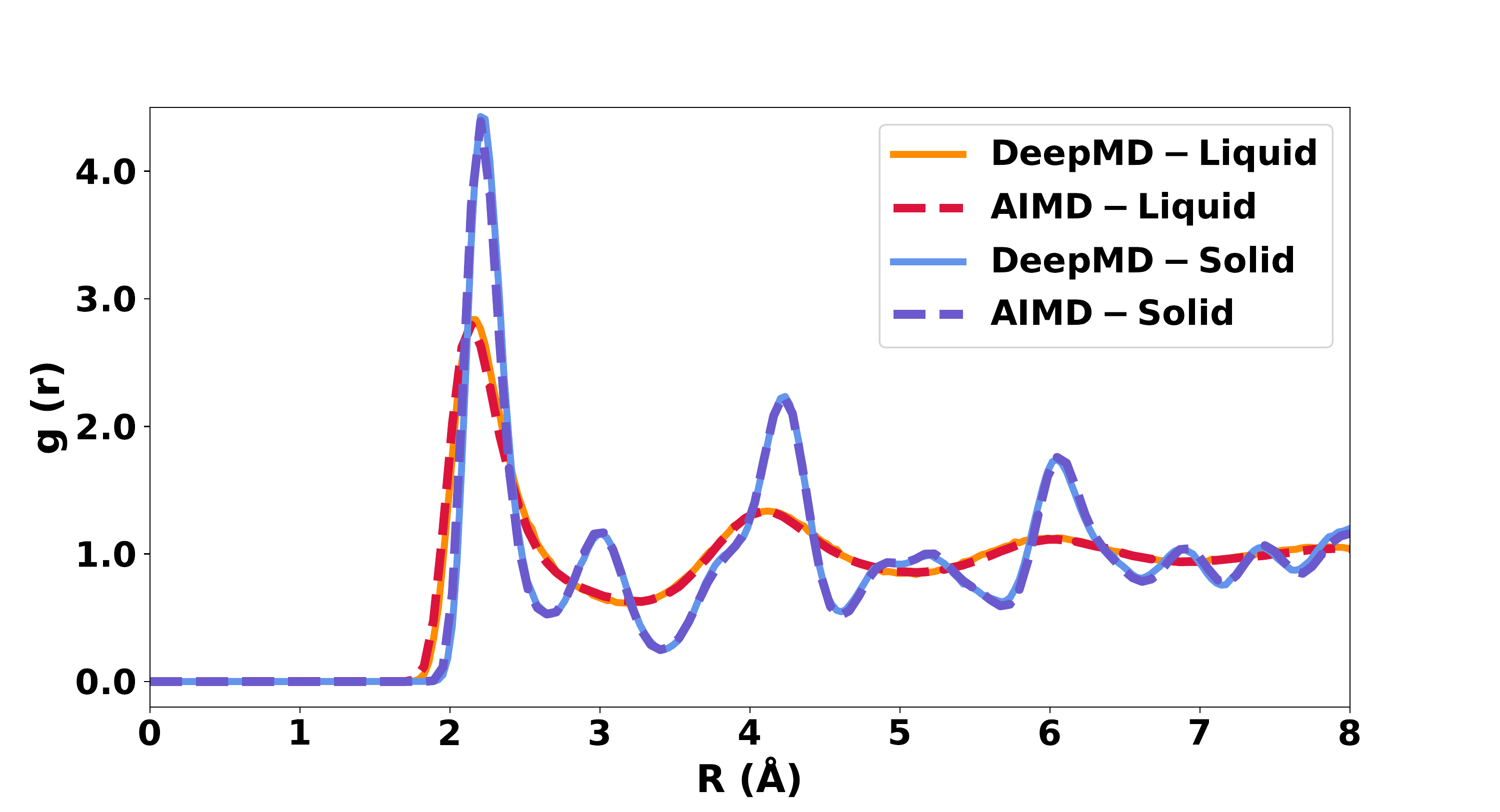}

\caption{(a) Root Mean Square Error versus data size of the training set. The errors are computed on a training set of 3k conﬁgurations which contains both data from the MetaD simulation and from equilibrated AIMD simulations. (b) Radial Distribution Function of the two phases, computed with the DeePMD potential. Reference data are from AIMD simulations.}
\label{e_f_accuracy}
\end{figure}

\begin{figure}[t]
\centering

\includegraphics[width=0.46\textwidth]{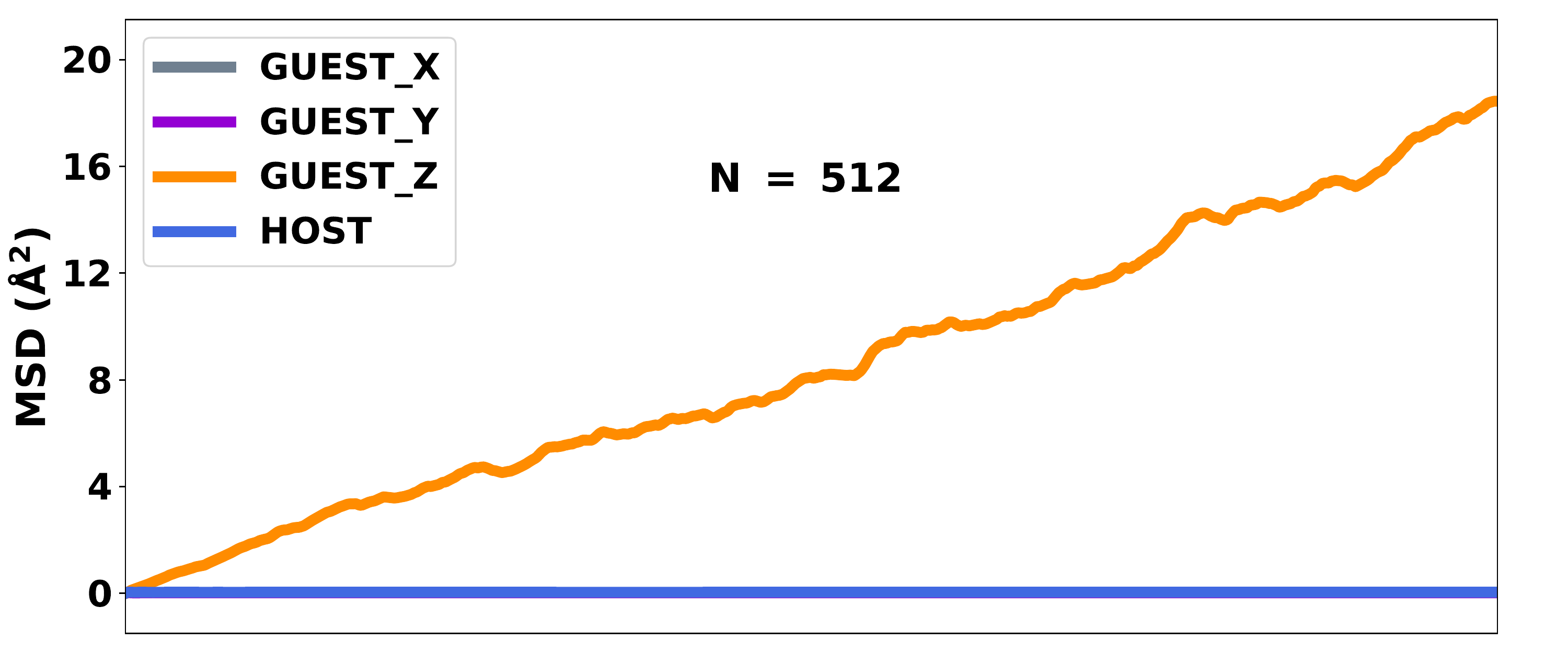}
\includegraphics[width=0.46\textwidth]{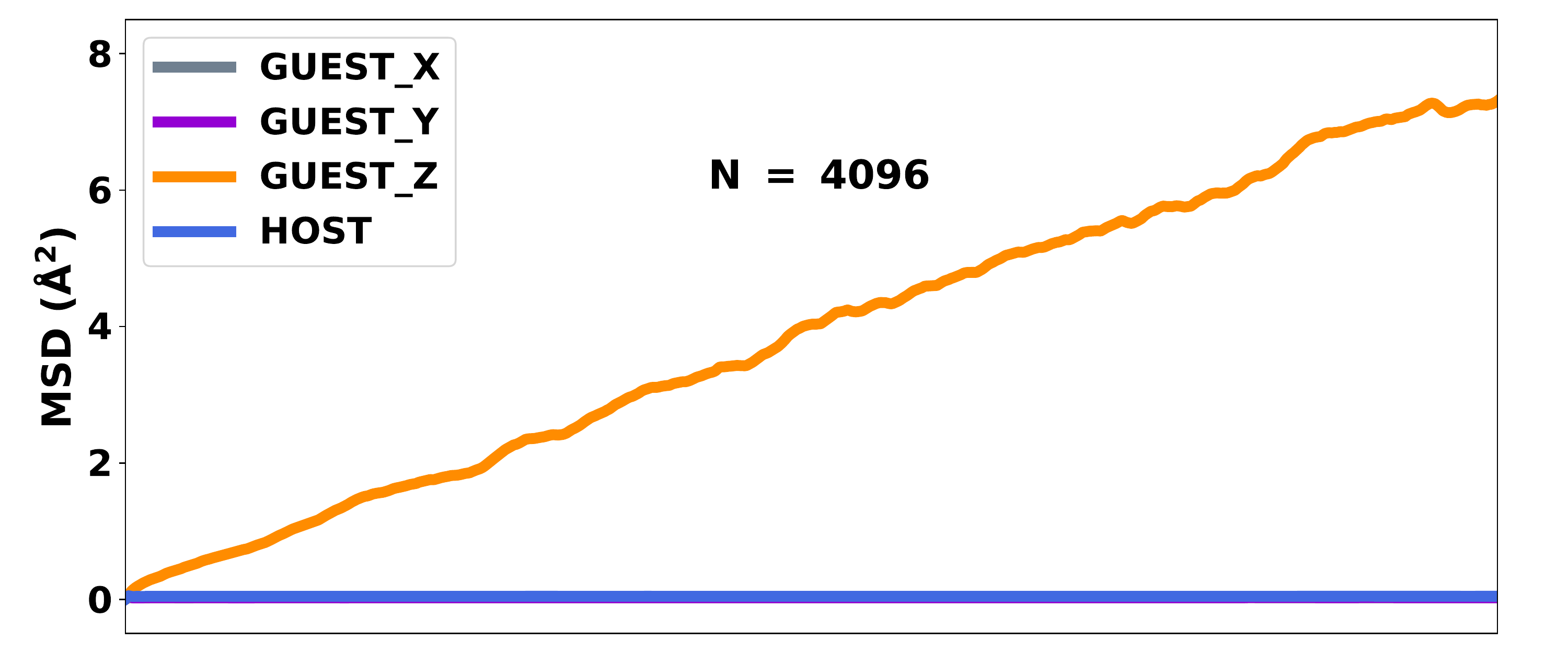}
\includegraphics[width=0.46\textwidth]{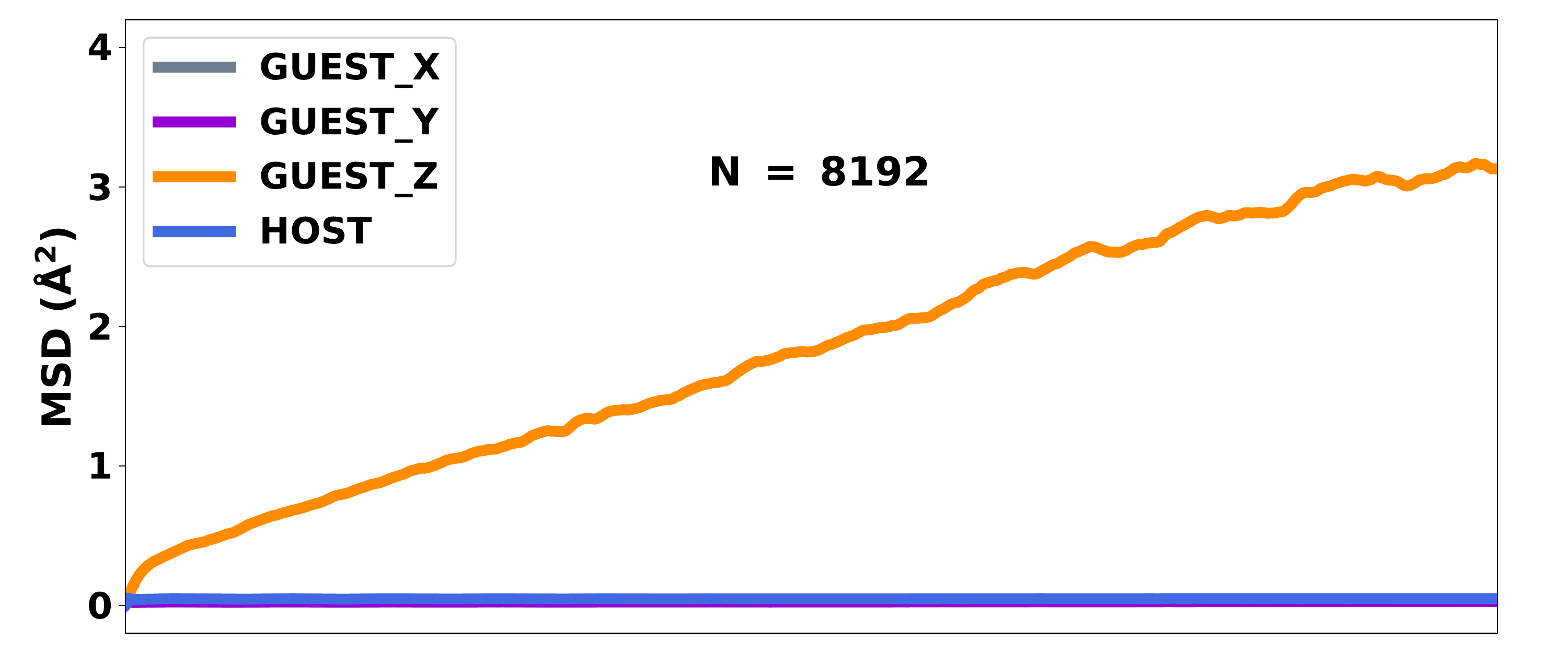}
\includegraphics[width=0.46\textwidth]{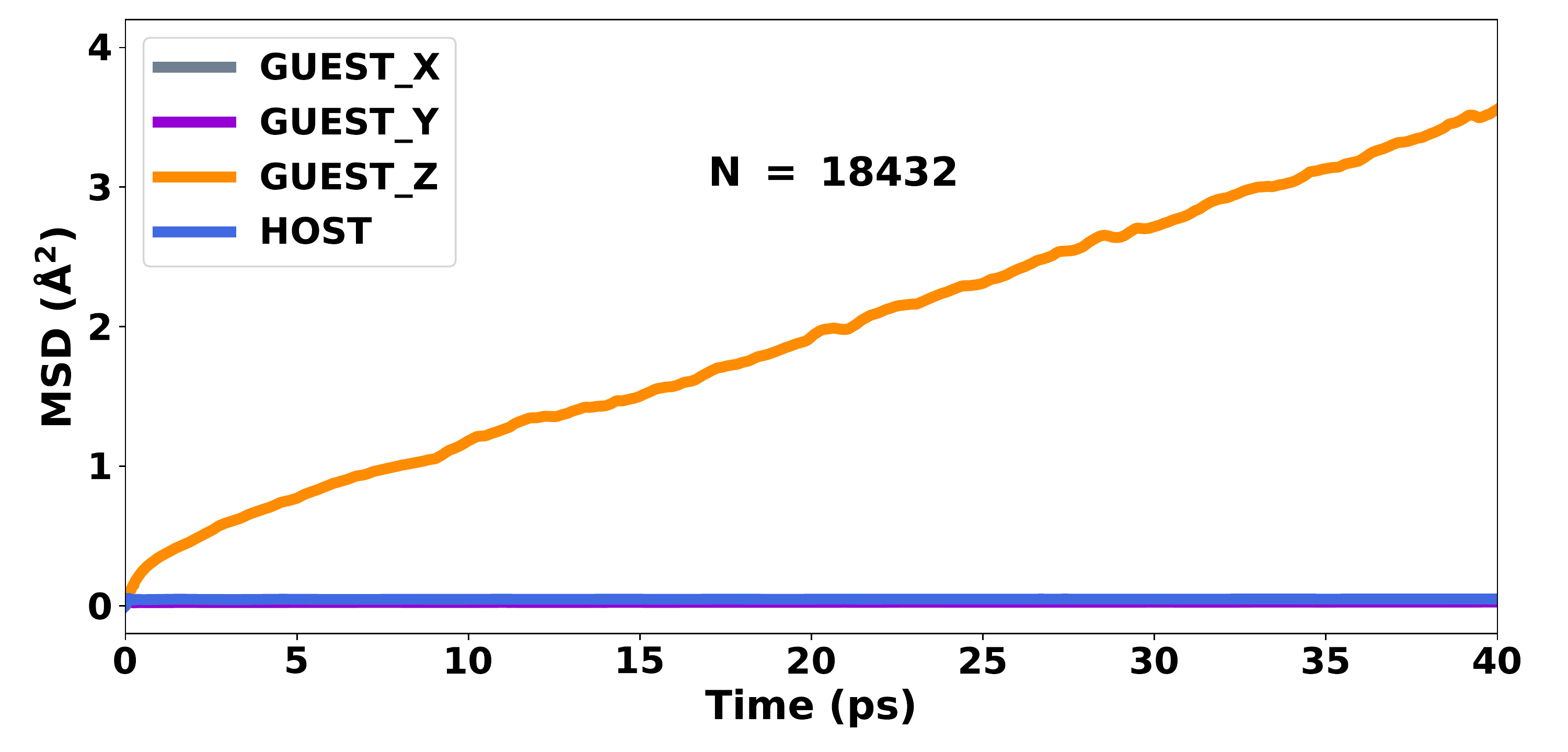}

\caption{(a) MSD from MLMD simulations, at P = 231 GPa, and T = 600 K and 1200 K respectively. We set the size of simulation cells N = 8192. (b) z-projected MSD of guest atoms from MLMD simulations versus increasing temperatures, at P = 231 GPa. We set the size of simulation cells N = 8192.}
\label{MSD}
\end{figure}

\begin{figure*}[t]
\centering

\includegraphics[width=0.46\textwidth]{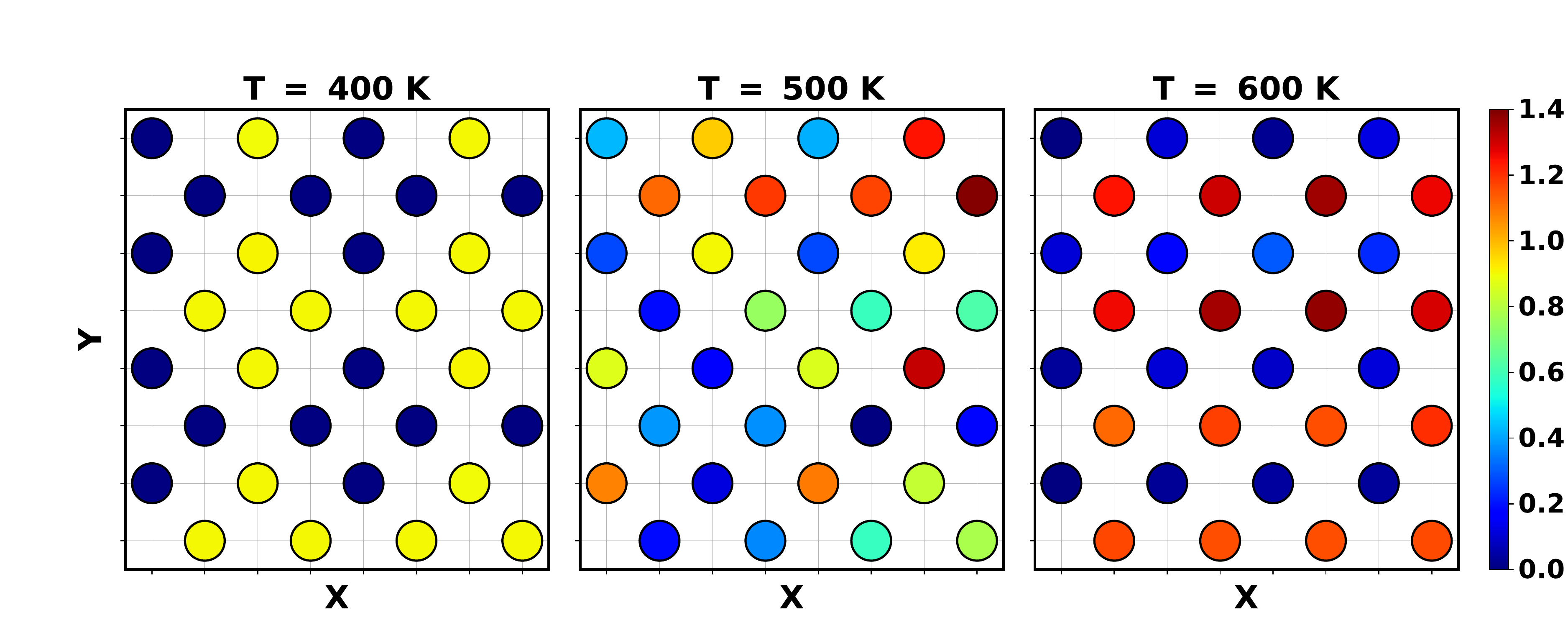}
\includegraphics[width=0.46\textwidth]{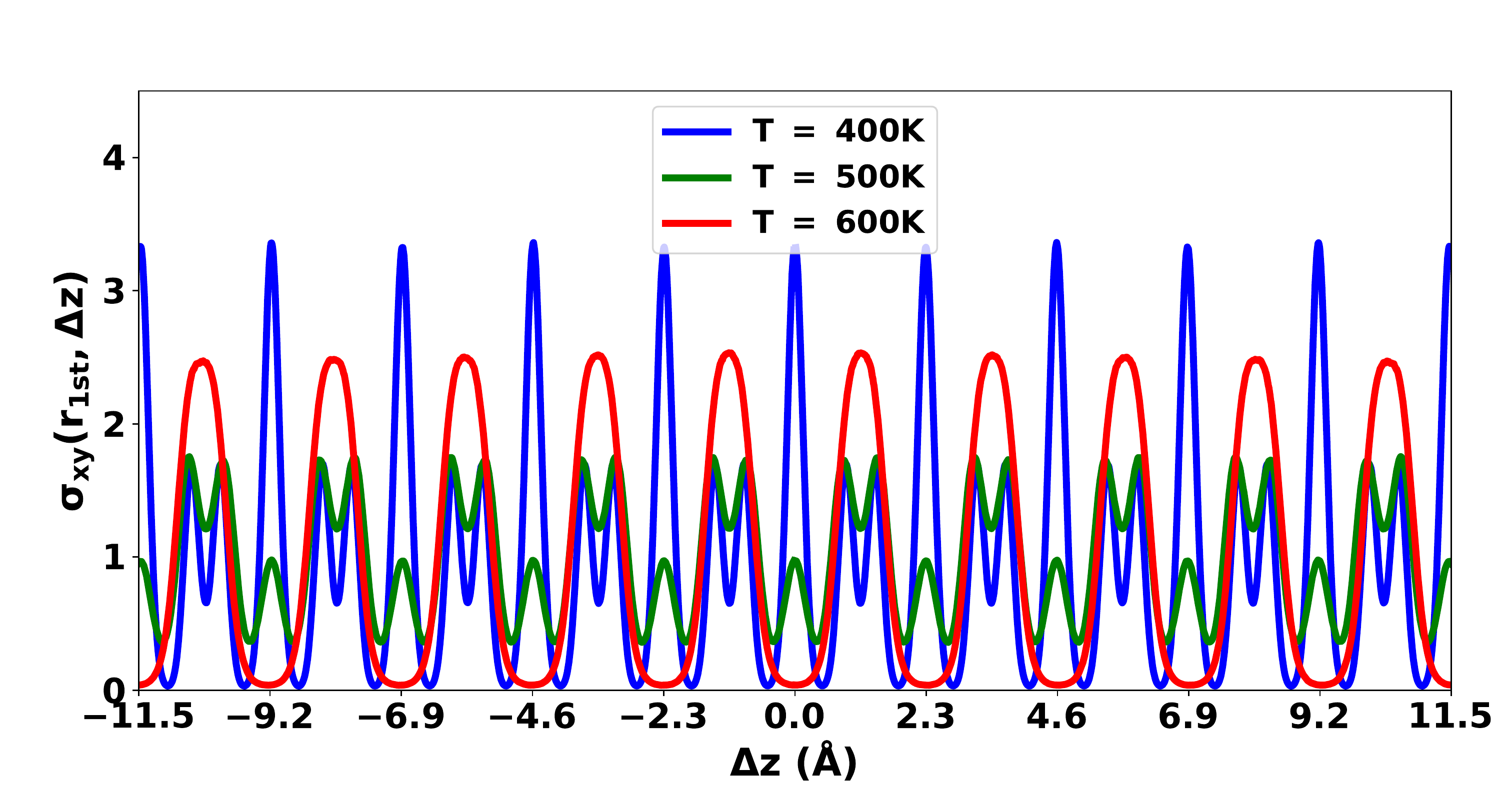}
\includegraphics[width=0.46\textwidth]{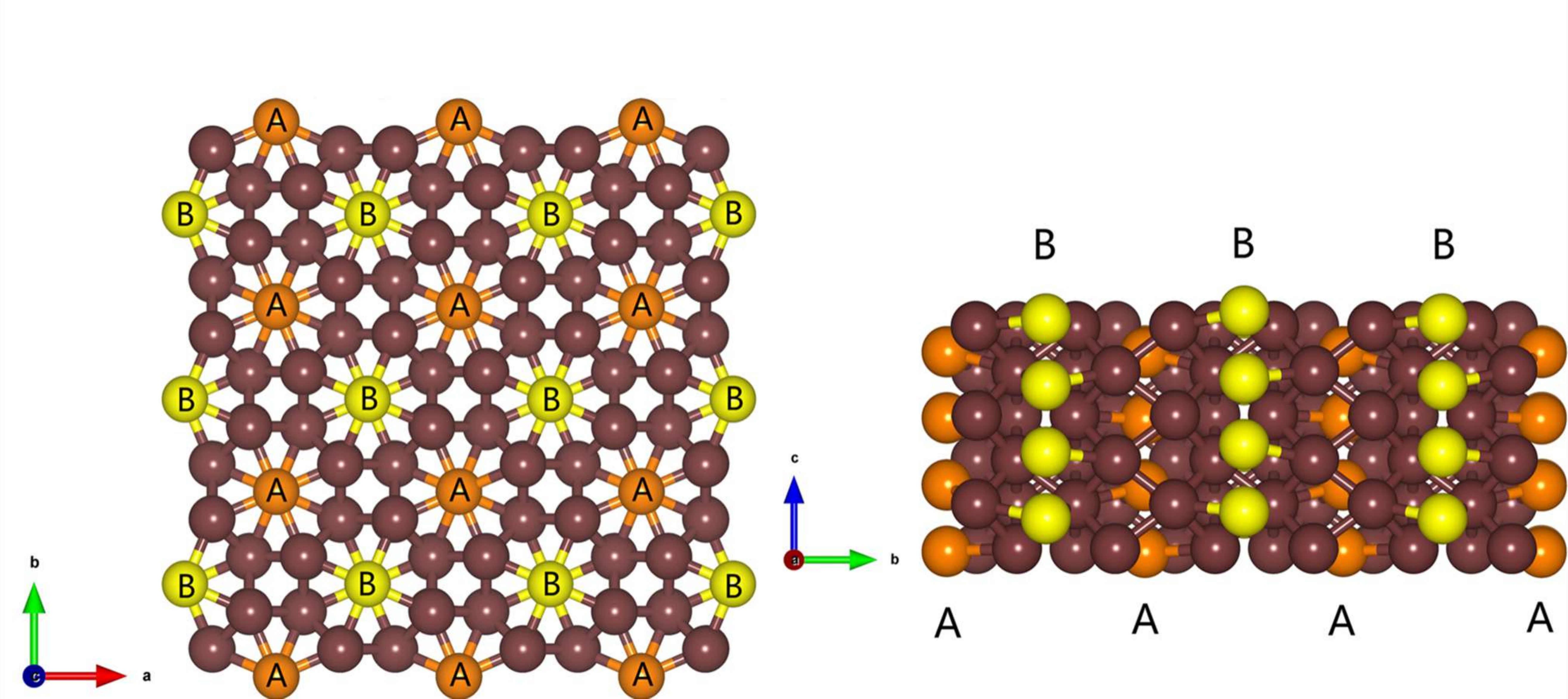}
\includegraphics[width=0.46\textwidth]{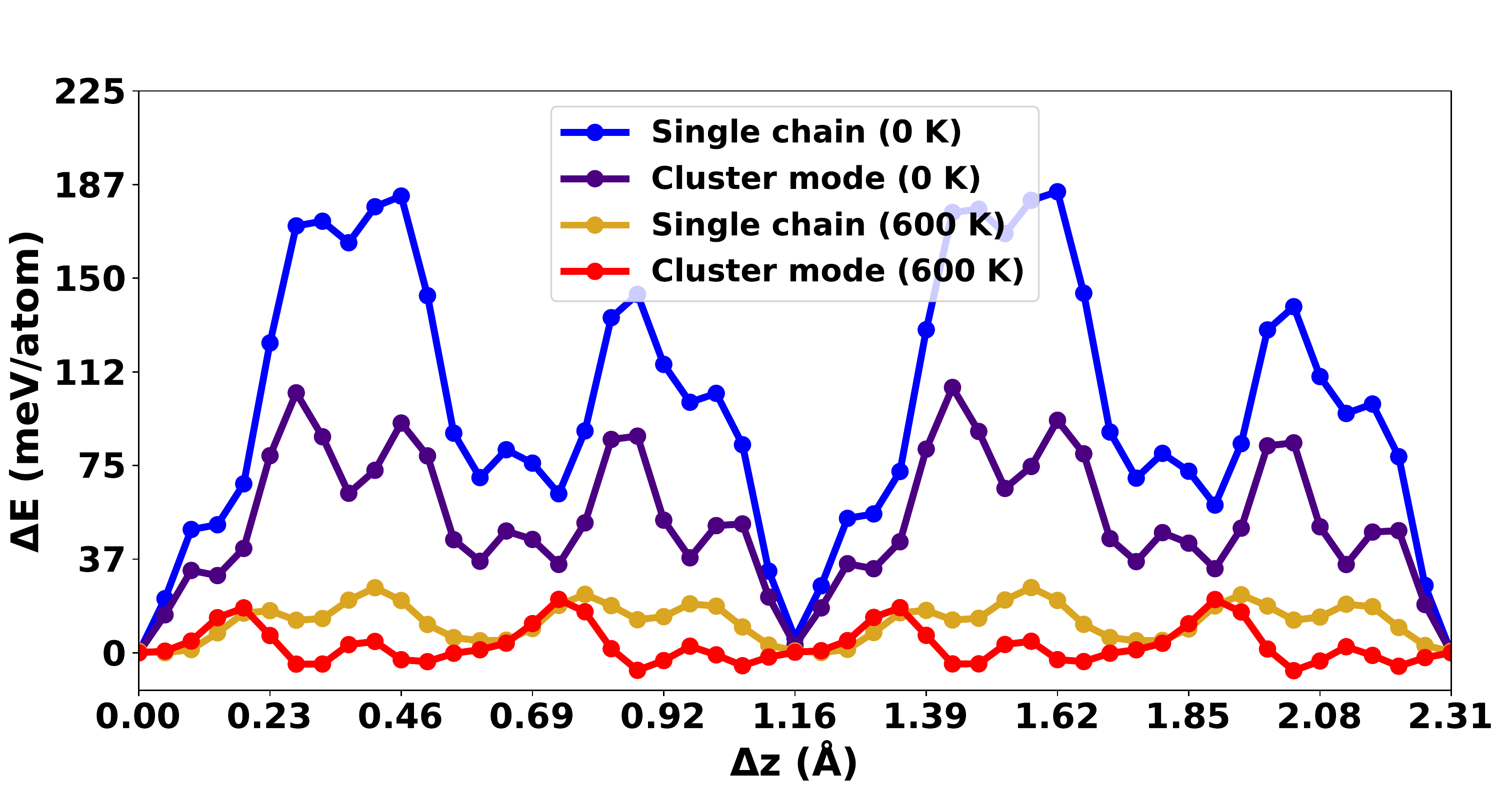}

\caption{(a) Average z sliding magnitude of guest chains at different temperatures, extracted from MLMD simulations. It clearly shows that the pattern of the guest chains transfers from a checkerboard type (400 K) to a stripe type (600 K). The colors represent the sliding amplitude of the guest chains in the unit of angstrom. (b) 1st-neighbour interchain correlation $\left<\sigma_{xy}(r_1,\Delta z)\right>$ at 231 GPa, at 400 K (blue), 500 K (green) and 600 K (red), respectively. (c) Average structure extracted from classical MD simulations (at T = 600 K and P = 231 GPa). (d) Energy barrier of sliding guest chain (average of each atom), calculated within the 0 K structure and newly emerged stripe type structure at 600 K, respectively.}
\label{chain_distribution}
\end{figure*}

\begin{figure}[t]
\centering

\includegraphics[width=0.46\textwidth]{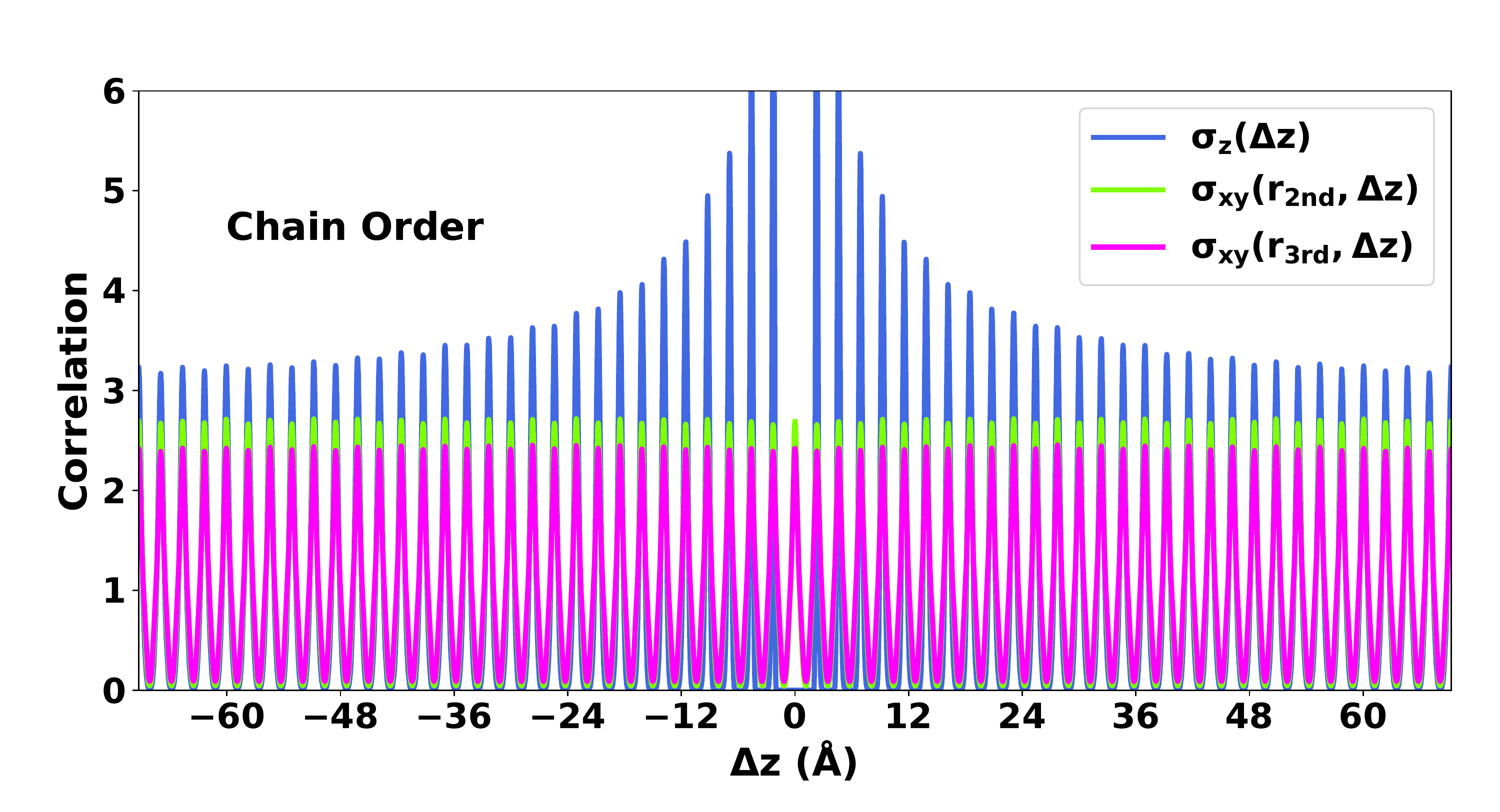}
\includegraphics[width=0.46\textwidth]{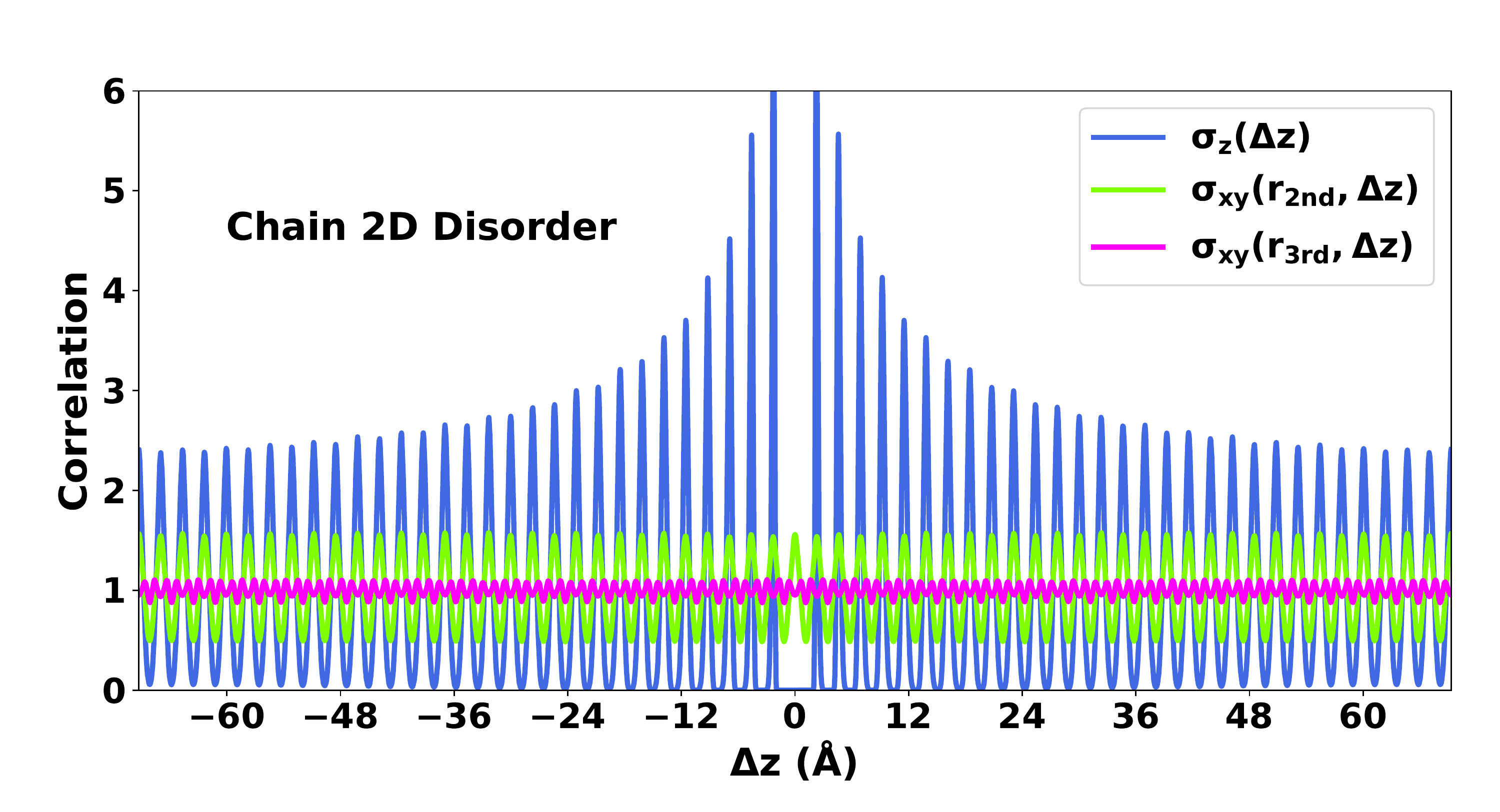}
\includegraphics[width=0.46\textwidth]{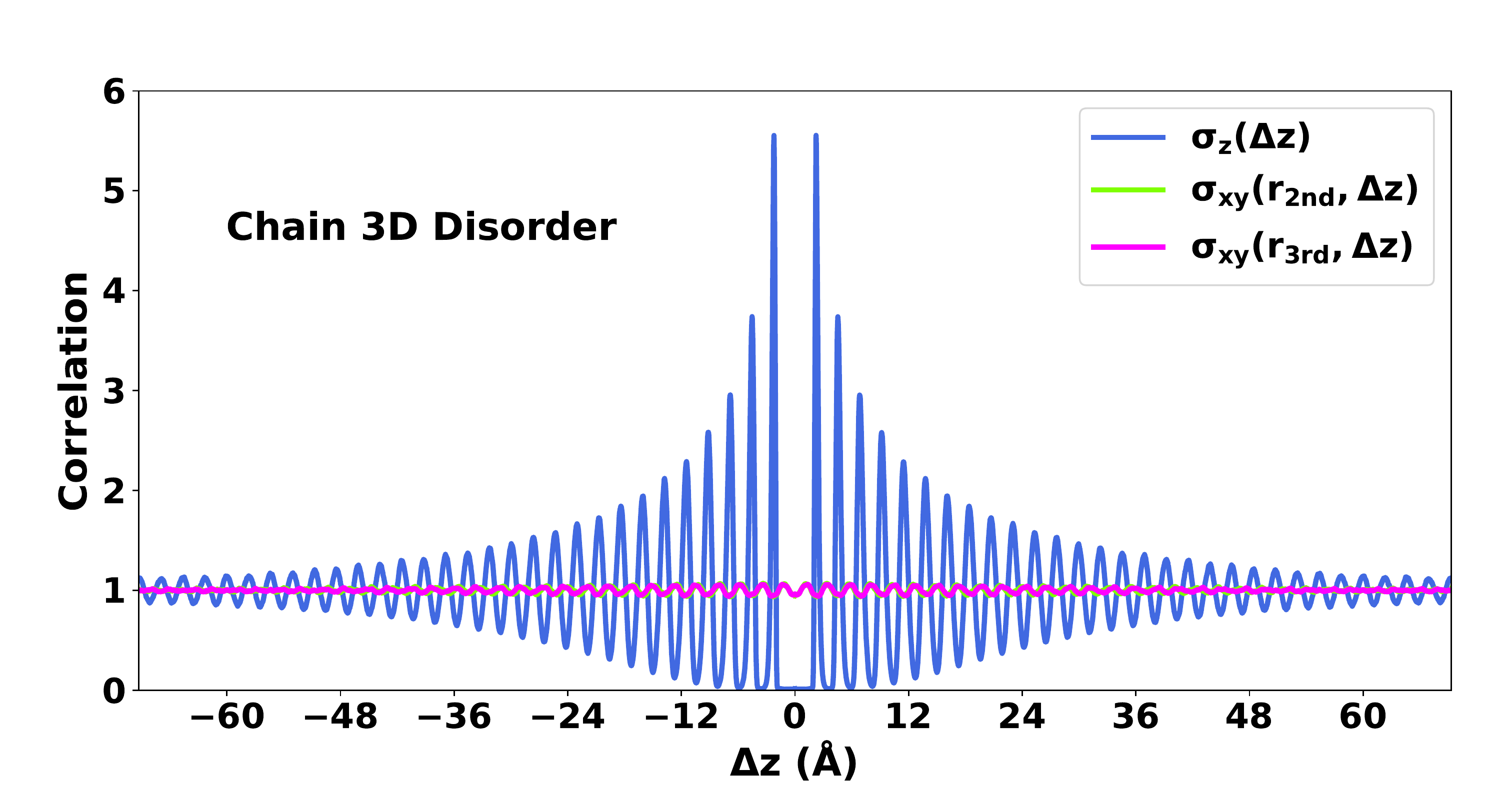}

\caption{Correlation $\sigma_z(\Delta z)$ along chain, and $\left<\sigma_{xy}(\Delta r,\Delta z)\right>$ for 2nd- and 3rd-neighbour chains. At 600 K, the guest sublattice keeps long-range order. At 900 K, the guest sublattice only loses inter-chain long-range order. At 1500 K, both intra- and inter-chain long-range orders disappear.}
\label{CORRELATION}
\end{figure}

\begin{figure}[t]
\centering

\includegraphics[width=0.46\textwidth]{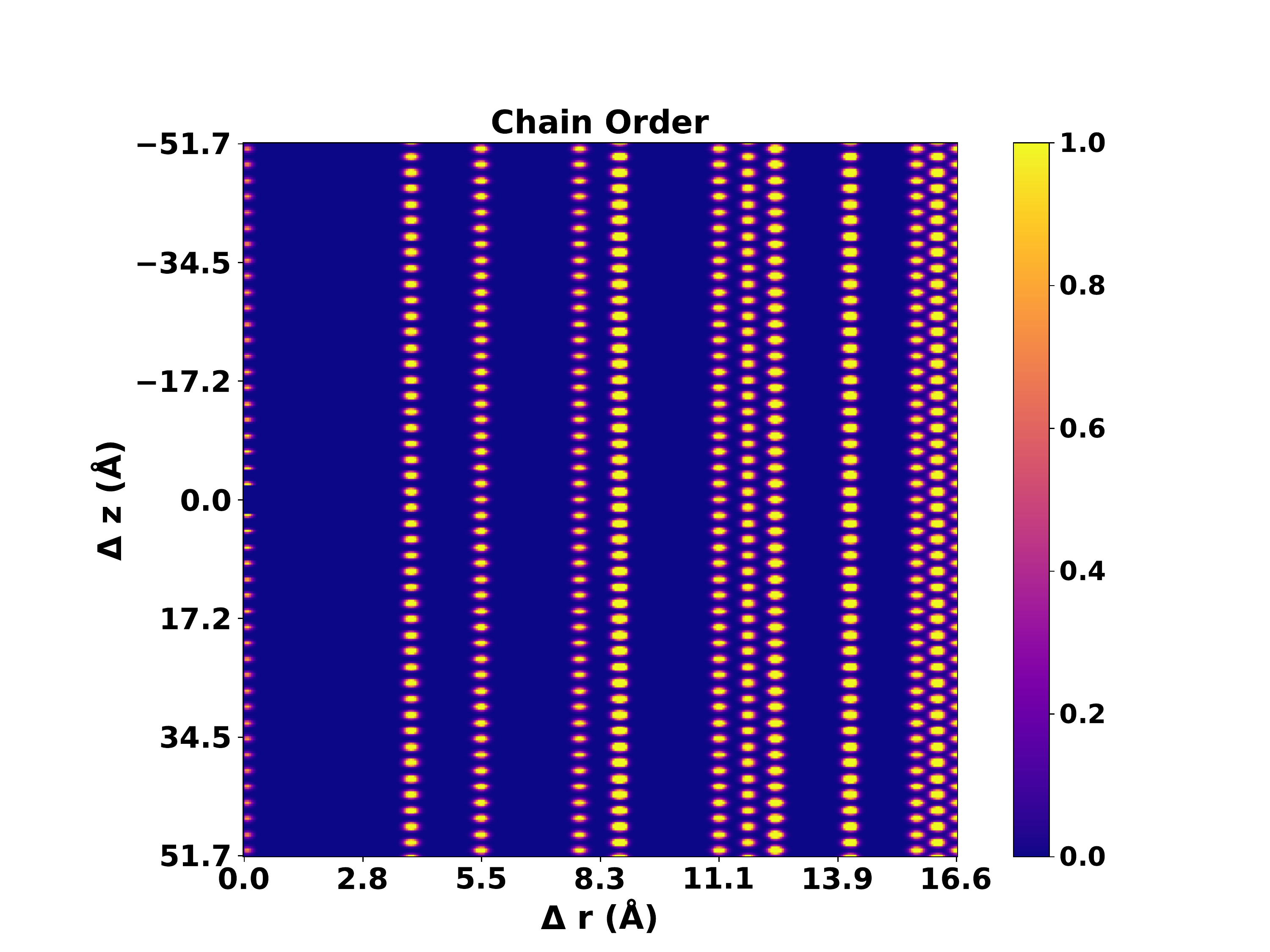}
\includegraphics[width=0.46\textwidth]{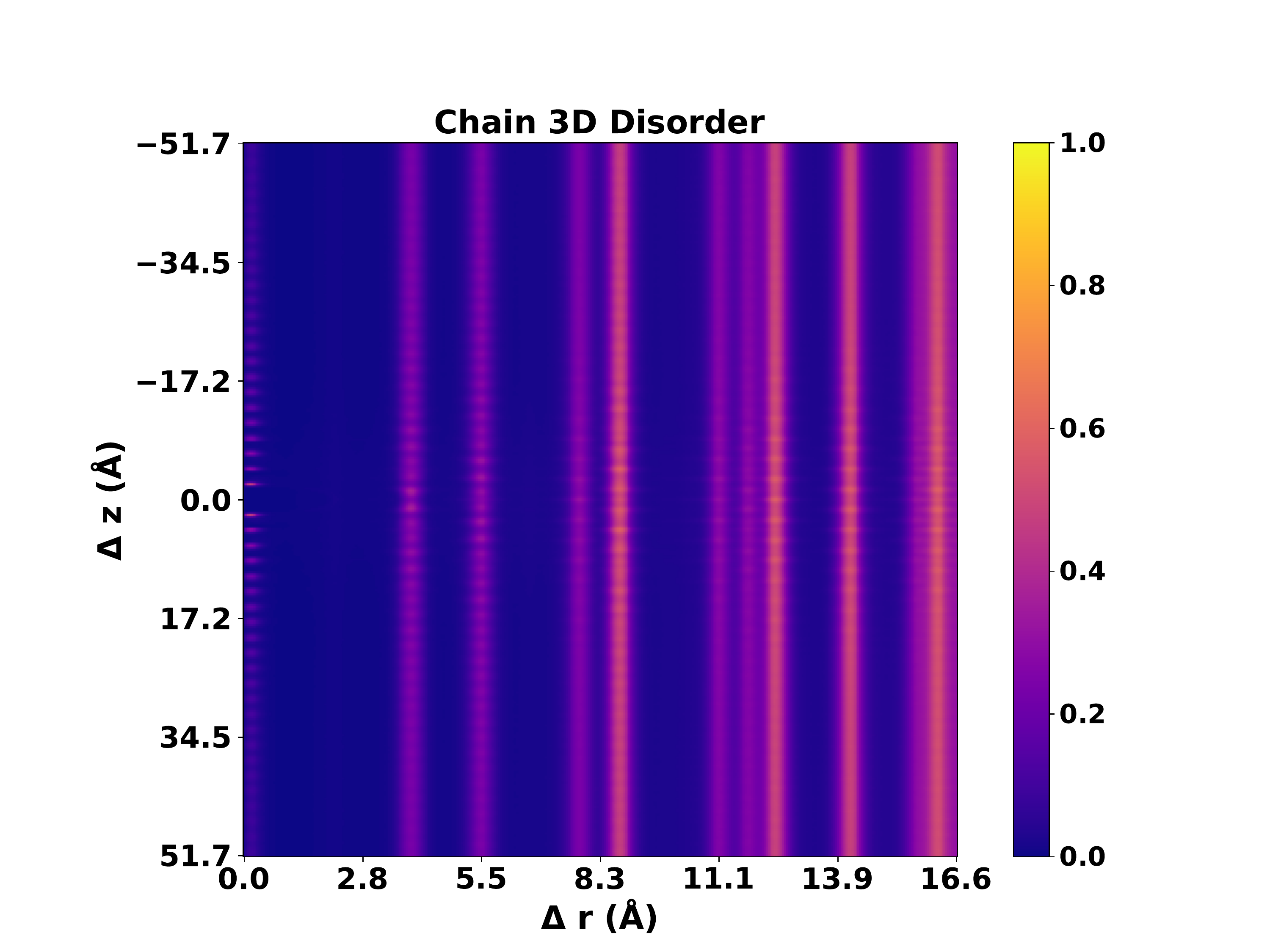}

\caption{(a) Normalized chain-correlation function $\sigma_{xy}(\Delta r,\Delta z)$ from MLMD simulations at 231 GPa, and T = 600 K (Chain Order) and 1500K (Chain 3D Disorder), respectively. $\Delta r$ and $\Delta z$ refer to separation in the $xy$ plane and along the $z$ axis respectively. See legends on the sides for color scheme, which range from $\sigma_{xy}=0$ (dark) to $\sigma_{xy}=1$ (white). Chain Order (T = 600 K) has well-defined peaks and thus long-range order both in-plane and along the chains.}
\label{2D_map}
\end{figure}

As shown in Fig.~\ref{vasp}~(a), the commensurate host-guest structure of calcium (Ca-VII) is composed of two sublattices. 1D (one-dimensional) atomic chains, which we call guest chains, are located in channels within the host sublattice (a zeolite-type structure). Distinguished by z coordinates, guest chains could be separated into two types, which we label A and B, as shown in Fig.~\ref{vasp}~(a). The guest atoms in the same chain does not form a straight line along z direction; instead, there are small transverse distortion respect to each other.

Firstly, we implemented AIMD simulations at 231 GPa and 1200 K. After long-enough simulation time, the arrangement of guest chains always exhibit a random distribution, consistent with the disappearance of the Bragg peaks of guest sublattice. To visualize the chain melting state, we extracted two snapshots from AIMD trajectories for comparison. As shown in Fig.~\ref{vasp}~(b), the first configuration is the input structure; while in the second one, we display the relative atomic displacements respect to input structure, which show that inter-chain correlations have disappeared.

Secondly, we calculated MSD (mean-squared displacement) to analyse quantitative behavior of the chain melting state. The divergence of MSD can indicate the melting state of a particular set of atoms in MD simulations. The MSD of a group of atoms is calculated using
\begin{equation}
\textup{MSD(t)}=\frac{1}{N}\sum_{i}\left|\textbf{r}_i(t)-\textbf{r}_i(0)-[\textbf{R}_{\textup{C}}(t)-\textbf{R}_{\textup{CM}}(0)]\right|^2
\end{equation}
where $\textbf{r}_i(t)$ is the position of atom i in the group at simulation time t, $\textbf{R}_{C}(t)$ is the center of mass of the group of atoms at t, and the summation runs over all atoms in the group, of which there are N atoms in total. We then calculated MSDs of guest and host atoms separately. At 600 K, MSD of both guest and host atoms are convergent through an enough long simulation time. Even though z-projected MSD of guest atoms shows a deviation from 0 initially, it is still recognized as typical solid-like oscillations versus time afterwards. While at 1200 K, z-projected MSD of guest atoms shows obvious divergence, growing nearly linearly with respect to time. The corresponding diffusion coefficient is approximately $1\times 10^{-4}cm^2/s$, consistent with a liquid phase. In contrast, x- or y-projected MSD of guest atoms and MSD of host atoms are always in the proximity of 0.

Then, with AIMD trajectories, we calculated phonon density of states by makeing Fourier transformation of VACF (velocity auto-correlation function), of guest and host atoms, respectively, as shown in Fig.~\ref{vasp_spectrum}. Zero-frequency compoent of guest atoms signal z-directional diffusion of guest atoms. Such diffusion can provide additional entropy to make the chain melting state thermodynamically stable. 2PT-MF (two-phase thermodynamic) model enables us to calculate both entropies and free energies for liquid metals, which is based on a decomposition of the velocity auto-correlation function into gas-like (diffusive) and solid-like (vibratory) subsystems~\cite{PRE-First-principles-calculation-of-entropy-for-liquid-metals}. We can analyse the entropy based on projected phonon density of states shown in Fig.~\ref{vasp_spectrum}. In the chain melting state, the entropy of host atoms is 1.88 kB/atom (per degree of freedom), and that of guest atoms projected on x or y direction is 1.78 kB/atom, while that on z direction is 2.47 kB/atom. Therefore, at 1200 K, the decrease of free energy induced by chain melting of guest atoms approaches 1.77 eV in a unit cell, providing a thermodynamical stabilization of the chain melting state.

The above evidence supports that the chain melting state is a phase in which guest chains are sliding freely along z-direction, without specific coherent manner. Besides, the state can be stabilized by the obvious decrease of free energy. But there could be severe finite size effects in AIMD simulations, which make the results unreliable. AIMD constrains us from further exploring the long-range order, since a simulation cell which is large enough is computationally demanding in AIMD simulations.

Therefore, we use the classical force field to run MD simulations, which can gauge the potential finite size effects in small supercells of AIMD simulations. We use the DeePMD (deep potential molecular dynamics) scheme~\cite{PRL-Deep-Potential-Molecular-Dynamics-A-Scalable-Model-with-the-Accuracy-of-Quantum-Mechanics,CCP-Deep-Potential,CCP-Deep-Potential} to construct the force field. DeePMD scheme has been successfully applied to simulate large-scale out-of-equilibrium system~\cite{PRL-Silicon-Liquid-Structure-and-Crystal-Nucleation-from-Ab-Initio-Deep-Metadynamics}. Even the NN potential is trained on a system of relatively small sizes (constrained by the simulation size in AIMD simulation), it can be used to study system of bigger sizes and long-range orders with DFT quality, due to the energy decomposition~\cite{PRL-Deep-Potential-Molecular-Dynamics-A-Scalable-Model-with-the-Accuracy-of-Quantum-Mechanics}. After defining the architecture of the neural network as described in Sec.~\ref{methods}, we optimized the parameters of the NN based on a training set. The energy converges smoothly as a function of the number of conﬁgurations included into the training set, as shown in Fig.~\ref{e_f_accuracy}~(a). We computed the RDF (radial distribution function) of both the solid (at 1200 K) and the liquid (at 3200 K) phases. An accurate description of the liquid state requires a discrimination of the metallic and covalent bonding, so the latter is an important benchmark for the quality of the force field. We compared RDF from AIMD simulation and that from classical MD simultions with DeePMD potential, and the result is remarkable, as shown in Fig.~\ref{e_f_accuracy}~(b).

After building the NN force field, we examined the convergence of MSD as a function of the size N of the simulation cell. Since the guest atoms diffuse along z direction, it's necessary to use a simulation cell with a long enough x axis, so as to obtain the correct quantitative behavior. Hence, rather than using cubic cells, we chose the simulation cells which are elongated in the c axis, including the 2*2*16 supercell (N = 8192) and 3*3*16 supercell (N = 18432). As shown in Fig.~\ref{MSD}, the MSD curves are not well converged until there are at least 8192 atoms in the simulation cell. We then calculated MSD versus temperature using a simulation cell with size N = 8192. It shows that the behavior of z-projected MSD of guest atoms can be separated into 3 stages as temperature increasing. Below 600 K, the diffusion coefficient of guest atoms along z direction is almost zero. At 800 K, its value quickly transits to $1\times 10^{-6}cm^2/s$. Above 1200 K, its values transits to $1\times 10^{-5}cm^2/s$. A major quantitative change is that the magnitude undergoes a nearly tenfold increase.

To better understand the decorrelation of guest chains in the chain melting state, we pay a particular focus on the loss of order within the guest sublattice. We define the chain correlation function $\sigma_{xy}(\Delta r,\Delta z)$ to detect the chain disorder within and between the guest chains, which gives
\begin{equation}
\begin{aligned}
\sigma_{xy}(\Delta r,\Delta z)=\left<\sum_{n,m}\sum_{i,j}\delta \left(z_{ni}-z_{nj}-\Delta z\right)\right. \\
\delta \left. \left(\sqrt{\left(x_{ni}-x_{mj}\right)^2+\left(y_{ni}-y_{mj}\right)^2}-\Delta r\right)\right>
\end{aligned}
\end{equation}
$n$ and $m$ are indices of the chains, and $i$ and $j$ label guest atoms within the $n-th$ and $m-th$ chain, respectively. For a solid guest lattice, $\sigma_{xy}$ shows long-range order along z and perpendicular direction. MLMD enables us to simulate chains of lengths up to 100 atoms. We calculate the correlation function based on the 18,432-atom Ca-VII supercell, with 72 chains and 64 atoms in each chain. Above the chain-melting transition temperature, we expect $\sigma_{xy}$ to detect loss of order between chains and along chains. We show normalized $\sigma_{xy}(\Delta r,\Delta z)$ from MLMD simulations at 600 K, 1200 K and 1800 K, respectively (see supplementary materials). For each simulation, the Ca sample is held at the selected temperature with NVT annealing for at least 20,000 steps. At 600 K, both host and guest structures remain solid. The guest lattice retains intrachain and interchain correlations and sharp peaks can be observed. MSD is convergent at this temperature. While at 1800 K, the peaks along z and along r have vanished in long range, indicating loss of correlation both within and between guest chains.

It is essential to seek clarification for the origin of chain melting transition. Usually, the energy barrier of sliding a chain relative to each other can be a small value in host-guest structures~\cite{PNAS-On-the-chain-melted-phase-of-matter}. We calculated the energy barrier of chain sliding of Ca-VII. For zero temperature structure, the energy barrier of sliding chains can approach as high as 150 meV (shown in Fig.~\ref{chain_distribution}~(d)), corresponding to 2000 K, quite close to the melting temperature of Ca-VII. Such a high energy barrier seems not to allows for 3D disorder of chains, in which every chain is able to slide freely against each other. We also calculated sliding energy barrier for a "cluster mode", in which we slide four neighbours in the same type (both sliding of four type A chains and type B chains give the same results). The maximum of energy barrier decreases to 100 meV, but it still corresponds to 1300 K, not consistent with starting temperature of chain melting, approximately 800 K. To explore the mechanism that reduces the sliding energy barrier, we extracted average z position of each guest chain, which reflects the order of guest sublattice. As shown in Fig.~\ref{chain_distribution}~(a), with the temperature increasing from 400 K to 600 K, the guest sublattice transits from zero temperature order to a new order. We calculated the 1st neighbour chain correlation function $\sigma_{xy}(r_1,\Delta z)$ to distinguish different orders of guest sublattice. As shown in Fig.~\ref{chain_distribution}~(b), at 400 K, peaks of $\sigma_{xy}(r_1,\Delta z)$ appear at at $z=n\cdot c_g$, $z=(n+\frac{1}{3})\cdot c_g$ and $z=(n+\frac{2}{3})\cdot c_g$, consistent with the structure of Ca-VII. While at 600 K, peaks of $\sigma_{xy}(r_1,\Delta z)$ transit to $z=(n+\frac{1}{2})\cdot c_g$. Additionally, 500 K appears to be an intermediate state, in which the peaks at $z=n\cdot c_g$ are weakened a lot, and the peaks at $z=(n+\frac{1}{3})\cdot c_g$ and $z=(n+\frac{2}{3})\cdot c_g$ start to incorporate to $z=(n+\frac{1}{2})\cdot c_g$. The new arrangement of guest sublattice at 600 K shows "ABAB..." pattern, different from the "AABB..." pattern at 0 K (as shown in Fig.~\ref{chain_distribution}~(c)). The long-range order transition of guest sublattice plays a significant role in lowering the energy barrier of chain sliding. As shown in Fig.~\ref{chain_distribution}~(d), we calculated the sliding energy barrier of the new structure emerging at 600 K, whose maximum is only 25 meV, corresponding to 300 K. In the new structure, the long-range order of guest sublattice would be easily destroyed; hence, it is natural to see that Ca-VII develops a 3D disorder state at a temperature much lower than melting temperature. Moreover, the " cluster mode" is also favorable in the 600 K structure. As shown in Fig.~\ref{chain_distribution}~(d), we bound five neighbour chains and move them together along z direction. The increase of energy (average value of each atom) is reduced compared to when we only move a single chain along z direction.

The next step is to examine whether the guest sublattice of Ca-VII develops a 3D disorder after the transition of guest-chain arrangement. We calculate intra-chain correlation $\sigma_{xy}(\Delta z)$ and inter-chain correlation $\sigma_{xy}(r_p,\Delta z)$ versus temperature from 600 K to 1800 K. We use $p=1$ and $p=2$ to reflect interchain correlations. The loss of order within the guest sublattice can be separated into three stages. As shown in Fig.~\ref{CORRELATION}~(a), at 600 K "chain order", the guest sublattice remains both intrachain and interchain correlations. At 900 K "chain 2D disorder", the guest sublattice loses the long-range interchain correlations. As shown in Fig.~\ref{CORRELATION}~(b), $\sigma_{z}(\Delta z)$ shows obvious peaks, while $\sigma_{xy}(r_{3rd},\Delta z)$ has become 1. Fluctuations of $\sigma_{xy}(r_{2nd},\Delta z)$ shows the guest sublattice only remains quite weak short-range interchain correlations. In this stage, guest chains are able to move against each other nearly freely, but atoms within one chain are still strongly correlated, moving with a coherent manner. Therefore, such a "chain 2D disorder" transition could also be interpreted as a transition to the incommensurate host-guest structure, which has a zero-frequency sliding mode in the phonon dispersion~\cite{SA-Strong-coupling-superconductivity-in-a-quasiperiodic-host-guest-structure}, in which the guest chains are able slide freely along c direction. This indicates that the guest sublattice may undergo another transition from commensurate HG structure to incommensurate HG structure around 800 K. The occurrence of such a two-dimensional disorder has already been observed in ${\mathrm{Hg}}_{3\ensuremath{-}\ensuremath{\delta}}\mathrm{As}{\mathrm{F}}_{6}$~\cite{PRL-One-Dimensional-Phonons-and-Phase-Ordering-Phase-Transition-in,PRL-One-Dimensional-Fluctuations-and-the-Chain-Ordering-Transformation-in}, in which a zero point in the phonon dispersion curve emerges when temperature increases above 150 K, corresponding to a free sliding mode between chains. At "chain 2D disorder" state, although the diffusion coefficient of guest atoms along z direction is divergent, its value is quite samll, approximately $1\times 10^{-6}cm^2/s$. When the temperature increases above 1200 K, we observed the guest sublattice transits to a "chain 3D disorder" state. From Fig.~\ref{CORRELATION}~(c) we can see that $\sigma_{z}(\Delta z)$ guest gradually decreases to zero, indicating that the sublattice loses intra-chain long range order. Therefore, both intra-chain and inter-chain long-range order disappear. The diffusion coefficient becomes 10 times larger than that in "chain 2D disorder" stage. Using extensive ab initio and machine learning molecular dynamics simulations, we find that the chain melting transition consists of several intermediate order phases. The guest sublattice firstly experiences a transition of long-range order arrangement, then 2D disorder and finally 3D disorder.

\bibliographystyle{apsrev4-1}
\bibliography{ARTICLE_main.bib}

\end{document}